# Older Adults' Safety and Security Online: A Post-Pandemic Exploration of Attitudes and Behaviors

Edgar Pacheco
*School of Information Management,*
*Victoria University of Wellington, New Zealand*
*edgar.pacheco@vuw.ac.nz*
0000-0003-4145-3244

**Abstract**
Older adults' growing use of the internet and related technologies, further accelerated by the COVID-19 pandemic, has prompted not only a critical examination of their behaviors and attitudes about online threats but also a greater understanding of the roles of specific characteristics within this population group. Based on survey data and using descriptive and inferential statistics, this empirical study delves into this matter. The behaviors and attitudes of a group of older adults aged 60 years and older (n=275) regarding different dimensions of online safety and cybersecurity are investigated. The results show that older adults report a discernible degree of concern about the security of their personal information. Despite the varied precautions taken, most of them do not know where to report online threats. What is more, regarding key demographics, the study found some significant differences in terms of gender and age group, but not disability status. This implies that older adults do not seem to constitute a homogeneous group when it comes to attitudes and behaviors regarding safety and security online. The study concludes that support systems should include older adults in the development of protective measures and acknowledge their diversity. The implications of the results are discussed and some directions for future research are proposed.

**Keywords** *internet usage; online risks, cybersecurity; online safety; digital literacy*

## 1. Introduction

Older adults' adoption and usage of digital technologies has significantly increased over the past decade (Faverio, 2022; Schehl et al., 2019) making them the population segment with the highest growth in this area (Friemel, 2016). The trend not only includes increased ownership of smartphones and tablet computers but also more frequent internet and social media use, as well as expansion of their online activities (Faverio, 2022; Schehl et al., 2019; Vaportzis et al., 2017). Evidence suggests that this tendency has been accelerated by the COVID-19 pandemic (Drazich et al., 2023; Lee, 2023; Sin et al., 2021; Sixsmith et al., 2022). However, while the trend depicts an improvement, adoption and usage of digital technologies continue to trail behind younger generations (Faverio, 2022; Pacheco, 2022) with evidence showing the gap extends to skills (Hargittai et al., 2019) and quality of internet connectivity (Pacheco, 2024).

Early academic literature has mostly portrayed older adults as unwilling or unable to use digital technologies (Vaportzis et al., 2017). However, this has progressively changed as benefits of digital tools perceived by older adults have been uncovered by research. These benefits include, for instance, communication with family and friends, access to information and services, leisure and recreation, dealing with social isolation and loneliness, maintaining independence, and higher life satisfaction and wellbeing (Aggarwal et al., 2020; Hou et al., 2022; Lips et al., 2020; Mohan & Lyons, 2022; Zhang et al., 2021). However, the growing adoption of digital technologies and the perceived advantages they offer to older adults is linked to the possibility of encountering online risks. The subsequent subsections delve deeper into this matter.

## 1.1 Older Adults and Online Safety and Security

Older adults' increasing engagement with and through digital technologies also means associated online risks and potential harm (Jiang et al., 2016). Online risks include a range of experiences, behaviors, and events such as cyberbullying, financial fraud, online harassment, identity theft, scams, and phishing. Livingstone (2013), who has extensively investigated online risks in relation to children, explains that conceptually online risks do not necessarily cause harm; however, when it occurs, harm involves negative impacts on people's emotional, physical, and/or mental wellbeing.

Another negative impact of online threats is financial loss, not only for those targeted but also the wider society (Moore, 2010). For instance, a 2023 survey of New Zealand's largest financial institutions reported that customers lost nearly $200 million NZD due to scams in the prior 12 months (RNZ, 2023).

Meanwhile, it has been argued that cybersecurity and online safety are two distinctive areas of policy and research (Robinson, 2022). The former, cybersecurity, involves protecting devices and networks, usually via technical measures (e.g., antivirus software or multi-factor authentication) from threats caused by third parties. Meanwhile, the latter, online safety, focuses on protecting the people using the devices and networks from threats and potential harm caused by third parties through awareness, education, information, and technology. However, while this distinction is helpful, in practice online safety and security can overlap. For instance, dealing with a privacy-related threat can involve technical and/or behavioral responses such as using antivirus software and/or stopping using the online service or tool.

Around the world, policy discussions and measures to reduce risks and possible harm from online threats have focused on improving people's online security (cybersecurity) and safety. In New Zealand, for instance, the Harmful Digital Communications Act 2015 (HDCA) serves as the primary legislative measure to address technology-mediated abuse. The HDCA, which is based on ten communications principles, introduced civil court orders for a serious and/or repeated breach of the communication principles and a criminal offense – sanctioned with fines and/or imprisonment – that makes it illegal to post a digital communication intended to cause harm or that in effect causes harm to the victim. Under the HDCA, an appointed Approved Agency receives complaints about harmful digital communications and promotes online safety (Pacheco & Melhuish, 2021).

## 1.2 The Extent of Online Threats among Older Adults

The extent of older adults' vulnerability to online threats has been a topic of research inquiry. A 2021 Pew Research Center study (Vogels, 2021) found that 21% of older adults aged 65 years and over experienced at least one form of online harassment (e.g., offensive name-calling. purposeful embarrassment, stalking, physical threats, harassment over a sustained period of time, and sexual harassment). By a large margin the most common experiences were offensive name-calling and purposeful embarrassment (Vogels, 2021). In terms of financial scams, another study shows that two thirds of older adults were targeted with COVID-19-related scams with charitable contributions and treatments being the most common (Teaster et al., 2023). Burnes et al. (2017) conducted a meta-analysis and found that financial fraud affects around 1 in 18 older adults (5.4%) annually, excluding those who are institutionalized or have cognitive impairments. Older adults are also the target of phishing attacks (Grilli et al., 2021) and consumer fraud (Shao et al., 2019) among others.

The literature has also compared and examined the attitudes and actions of older adults regarding online security and safety. It is argued that, compared to their younger counterparts, older adults are more concerned about their privacy and security online (Branley-Bell et al., 2022); however, they tend to be more reluctant to ask for help or to report that they have been the target of an online threat (Parti, 2022). They are also less aware and knowledgeable about managing security and privacy risks such as sharing passwords, using the same password over multiple platforms or passwords that can be easily guessed (Grimes et al., 2010). When it comes to participating in work or school-based safety training, older adults are less likely to do so (Alagood et al., 2023).

What is more, it is argued that as older adults tend to have lower digital literacy and skills, which have an impact on their protective behaviors online (Schreurs et al., 2017). When compared with their younger counterparts, older adults are less likely to use password managers (Ray et al., 2020) and report low adoption of two-factor authentication systems (Das et al., 2021). In addition, having less confidence with digital tools results in older adults performing fewer security behaviors, experiencing difficulties using security tools, spotting risks, and making informed privacy decisions (Jiang et al., 2016; Morrison et al., 2021). It has also been argued that due to declining cognitive abilities, they struggle to identify fake emails and websites (Mentis et al., 2019).

To sum up, there is a growing body of research about older adults' experiences of online threats. However, with the continuous evolution of the internet and related technologies, new online threats may arise, posing potential risks and harm. Therefore, it is crucial to continue gathering evidence on this phenomenon to identify patterns and its impact on vulnerable groups such as older adults.

### 1.3 The Heterogeneity of Older Adults

News outlets and social media platforms have consistently described older adults as a homogenous group, portraying them as disempowered, vulnerable, and passive (Makita et al., 2021). This has also been a tendency in research on internet usage, skills, and inequalities, in which older adults have been overlooked and described as a group with identical online behaviors and experiences (Hargittai & Dobransky, 2017; Quan-Haase et al., 2018). This line of thought and operationalization of older adults in research has been found in the studies summarized so far in this section. The risk of assuming older adults are a uniform group is that this view undervalues their agency to control different facets of their daily lives and ignores that they carry with them the skills and attitudes they have developed growing up (Quan-Haase et al., 2018). What is more, as Mannheim et al. (2019) point out, the homogenization of older adults can not only perpetuate ageist assumptions, but also hinder the development of technologies that meet their diverse needs, excluding them from research and design processes.

However, over the last decade a number of studies have provided evidence of the heterogeneity of older adults when it comes to their adoption and usage of the internet. A nationally representative survey (van Deursen & Helsper, 2015), for instance, found that older adults over 75 exhibit less variety in their internet use and are less likely to use email than older adults aged between 65 and 70. Additionally, older men use the internet more frequently than older women (van Deursen & Helsper, 2015). Another quantitative study (Hargittai & Dobransky, 2017) showed that older adults with greater incomes and educational levels are better at using websites, and that they are also more likely to utilize the internet for a wider range of purposes. A qualitative study with a group of Canadians aged 65 and above (Quan-Haase et al., 2018) similarly found that although they reported limited internet skills, they engaged in a wide range of online activities such as keeping in touch with family and friends, searching for information, performing economic activities (e.g., online banking), and entertainment (e.g., watching YouTube). Meanwhile, Liu (2021) observed that within the subset of regular internet users who possess a favorable social status, white females experienced a higher level of well-being and a diminished sense of loneliness. Conversely, black and minority ethnic females with an unfavorable social standing, despite their consistent use of the internet, reported a lower quality of life and a heightened feeling of loneliness. In another study conducted by Medero, Merrill Jr., and Ross (2022), it was found that older adults from racial and ethnic minority backgrounds demonstrated a decreased frequency of internet usage on both personal and public computers. Meanwhile, in Finland, Hänninen et al. (2021) conducted a qualitative study with adults aged 75 and above. The study found that participants' technology engagement varied: some were competent users, others avoided any engagement with the digital environment due to lack of interest, and some others engaged with the support of knowable users.

To sum up, there is an emerging literature about variations in internet access and activities within the segment of older adults showing they represent a heterogeneous group. However, to the best of our knowledge, minimal scholarly attention has been devoted to whether differences are also found in terms of cybersecurity and online safety behaviors and experiences. This is indeed surprising, given the susceptibility of this demographic to various forms of online threats, including scams and financial fraud. Consequently, the current work seeks to build on existing efforts by providing a snapshot of evidence from New Zealand. In doing so, policies and services can be designed and implemented to promote a safer online environment for older adults.

## 2. Method

### 2.1. Research Aims

The aim of the current work is to explore the attitudes and behaviors of older adults from New Zealand regarding some dimensions related to online safety and cybersecurity, and whether differences based on key demographics are statistically significant. The dimensions of interest are the following: attitudes towards the security of personal information online, decisions to stop using an online service due to privacy concerns, common protective actions against online threats, awareness of where to report an online threat, and attitudes towards six online safety issues (i.e., cyberbullying, online extremism, identity theft, misleading information, conspiracy theories, and hate speech). The current work seeks to address two core research questions (RQ):

**RQ 1:** What are the attitudes and behaviors of older adults aged 60 years and above from New Zealand about different dimensions related to online safety and security?

**RQ 2:** Are there significant differences in older adults' online safety and security attitudes and behaviors in terms of gender, age group, and disability status?

## 2.2. Study Design

The data come from *New Zealand's Internet Insights*, a survey aiming at gathering evidence about different aspects of adult New Zealanders' interactions with digital technologies. Secondary analysis was conducted on the data related to online security and safety behaviors and attitudes.

The survey was administered online. Online surveys are cost-effective, easier to administer, and increasingly used in social and policy research (Lehdonvirta et al., 2021). In terms of exploratory inquiry, online surveys have been shown to be a useful technique for gathering evidence about people's self-reported access, activities, and attitudes regarding the digital environment as well as experiences of online victimization (Pacheco & Melhuish, 2018, 2019; Pacheco & Burgess, 2024).

Data collection was conducted from November 7 to 14, 2022, by a market research company on behalf of InternetNZ, a not-for-profit organisation managing the .nz country code top-level internet domain. Participants were recruited from the company's online consumer panels. A combination of pre-survey quotas as well as post-survey weighting based on census estimates from Stats NZ (New Zealand's official data agency) were employed to ensure the overall sample is representative of key demographics such as age and gender.

The original full survey dataset included adult internet users aged 18 years and above (n=1,001). A total of 275 participants reported to be aged 60 years and older. This group encompassed the final sample for the current study. Similar to other studies on older adults and digital technologies (Hargittai et al., 2019; Jacobson et al., 2017), this age criterion was chosen to focus specifically on participants of the older adult demographic.

The survey questionnaire underwent cognitive testing and pilot testing.

When data collection took place, the COVID-19 Protection Framework had already been terminated by the New Zealand government. The framework was put in place to make life with the Omicron variant easier to handle. It operated from December 2021 to September 2022. Due to declining reported COVID-19 cases, a highly immunized population, and improved availability of antiviral medications to treat COVID-19, the rules for various "traffic light" settings were eliminated. A New Zealand Government website provides more detailed information about the framework (New Zealand Government, 2022).

The maximum margin of error for the total sample is +3.1% at the 95% confidence interval.

## 2.3. Ethical Considerations

In relation to ethical considerations, all participants provided informed consent, which was documented. They were briefed on the survey's purpose, its topics, and how their information would be utilized. Additionally, the survey introduction outlined participants' right to abstain from the research and to refuse to answer any questions. It also assured them the right to withdraw from the survey at any point. Furthermore, participants were informed about data storage. The market research firm responsible for data collection ensured that participants' personal information would not be linked to their responses, thus guaranteeing confidentiality. During the data collection phase, the market research company adhered to the New Zealand Research Association's Code of Ethics.

## 2.4. Measures

### 2.4.1. Demographics

To account for demographic characteristics associated with experiences and attitudes regarding online safety and security, we included gender, age group, and disability status as independent variables. Of the sample, 46.9% identified as male and 53.1% identified as female. No participant in our sample of older adults identified as gender diverse. Age group was a dichotomous variable with participants aged 60-69 representing 52.4% of the sample while participants aged 70 and above represented 47.6%. In terms of disability status, 76.3% indicated not having a disability while the remaining 23.7% said they do. The definition of disability used in the 2013 Disability Survey (Stats NZ, 2014) was applied to identified participants who self-reported whether they experience a disability. In this respect, participants were asked: 'Do you have a long-term disability or impairment (lasting six months or more) that makes it more difficult for you to do everyday tasks, that other people find easy?' Participants who responded 'Don't know' or 'Prefer not to say' were excluded from analysis (n=9).

### 2.4.2. Online safety and security

The survey included questions regarding experiences and attitudes about different aspects of online safety and security. In this respect, participants were asked how concerned they were about the security of their own personal details when they use them on the internet. The possible responses were: 'Extremely concerned', 'Very concerned', 'A little bit concerned', 'Not very concerned', 'Not at all concerned', and 'I don't use personal details on the internet'.

Then, participants were asked the following question: 'In the last 12 months, have you decided not to use an online service because of security concerns?' The following options were provided to answer the question: 'Yes – many times', 'Yes – one or just a couple of times', and 'No'.

Then, participants were asked about what security precautions they take with their own devices. This was a multiple-choice question which listed the following options: 'Use a PIN or password on any or all devices', 'Regularly update the software', 'Regularly back up the content', 'Use two factor or multi factor authentication on any or all accounts', 'Use a password manager', 'Use unique passwords for every online service', 'Use a VPN if connecting to the internet using a Wi-Fi network you don't know', and 'None of these'. Only descriptive analysis was performed on the data provided for this question.

Regarding online safety, the following question was asked: 'If you see something online that might be concerning, harmful or dangerous, do you know where to report it?' It was clarified to participants that this included websites, organizations, or agencies to go to. The response options for this question were: 'Yes', 'No', and 'Don't know'.

To measure attitudes about online risks, Likert-type items were used to ask participants how concerned they were about cyberbullying, online extremism, identity theft, misleading information, conspiracy theories, and hate speech. For each of these risks listed, participants were asked to think about society instead of anything they may or may not have personally experienced. To rank each of these online risks, the following scale was provided: 'Extremely concerned', 'Very concerned', 'A little bit concerned', 'Not very concerned', 'Not at all concerned', and 'Don't know'. Participants who selected 'Don't know' were excluded from analysis.

Finally, the survey included a question about attitudes towards the internet. Participants could choose between 'Yes', 'No', and 'I don't know' to answer the following question: 'There are positives and negatives to the internet, but overall do you think the positives outweigh the negatives?'. Participants who selected 'Don't know' as an answer were excluded from analysis.

### 2.4. Analytical Approach

All analyses were performed using the Jamovi program, version 2.3 (The Jamovi Project, 2023). The analysis comprised a dual framework, employing both descriptive and inferential statistical methods. Descriptive statistics were used to provide an initial overview of the dataset, while inferential statistics, specifically the Chi-square test of independence, were utilized to delve deeper into potential associations between categorical variables. Statistical significance was determined at $p < 0.05$. Simultaneously, we extended our inferential framework by incorporating Cramer's V statistic. Cramer's V measures the strength of association between categorical variables. A Cramer's V below 0.10 means a negligible association, between 0.10 and 0.20 indicates a weak association, between 0.20 and 0.40 represents a moderate association, and between 0.40 and 0.60 indicates a relatively strong association (Rea & Parker, 2014). Categories regarding online security and level of concern about online risks were collapsed into larger categories to facilitate analysis of the observed counts.

## 3. Results

To begin with, the association of the level of concern about the security of one's own personal details when using the internet with each of the three demographic variables was tested. As can be seen in Table 1, there was a significant association between level of concern about personal details online and age, $\chi^2$ (3, N = 275) = 8.40, p = .038. However, the effect size for this finding, Cramer's V, was weak: .17. Participants aged 70 and over were slightly more likely to be extremely/very concerned about the security of their personal details than those aged 60-69 years old. On the other hand, no significant associations were found in terms of gender, $\chi^2$ (3, N = 275) = 5.59, p = .133, and disability status, $\chi^2$ (3, N = 266) = 1.12, p = .772.

**Table 1. Participants' Attitudes about security of Personal Details by Gender, Age Group, and Disability Status**

| | Security of personal details when using the internet | | | | | | | | | | |
|---|---|---|---|---|---|---|---|---|---|---|---|
| | Extremely/very concerned | | A little bit concerned | | Not very/not at all concerned | | I don't use personal details online | | Total | | *p* |
| | n | % | n | % | n | % | n | % | n | % | |
| Gender | | | | | | | | | | | .133 |
| Male | 81 | 62.8 | 38 | 29.5 | 5 | 3.9 | 5 | 3.9 | 129 | 100.0 | |
| Female | 72 | 49.3 | 62 | 42.5 | 5 | 3.4 | 7 | 4.8 | 146 | 100.0 | |
| Total | 153 | 55.6 | 100 | 36.4 | 10 | 3.6 | 13 | 4.7 | 275 | 100.0 | |
| Age group | | | | | | | | | | | .038 |
| 60-69 | 78 | 54.2 | 60 | 41.7 | 2 | 1.4 | 4 | 2.8 | 144 | 100.0 | |
| 70+ | 75 | 57.3 | 40 | 30.5 | 8 | 6.1 | 8 | 6.1 | 131 | 100.0 | |
| Total | 153 | 55.6 | 100 | 36.4 | 9 | 3.6 | 12 | 4.4 | 275 | 100.0 | |
| Disability status | | | | | | | | | | | .772 |
| With disability | 38 | 60.3 | 21 | 33.3 | 2 | 3.2 | 2 | 3.2 | 63 | 100.0 | |
| Without disability | 108 | 53.2 | 77 | 37.9 | 8 | 3.9 | 10 | 4.9 | 203 | 100.0 | |
| Total | 146 | 54.9 | 98 | 36.8 | 10 | 3.8 | 12 | 4.5 | 266 | 100.0 | |

In terms of deciding not to use an online service in the past year due to security or privacy concerns, just over 60% of participants indicated they stopped doing so. However, none of the three demographic variables tested were associated with participants' behavior. The Chi-square test of independence showed the following results: gender, $\chi^2$ (1, N = 275) = 0.154, p = .695, age, $\chi^2$ (1, N = 275) = 0.510, p = .475, and disability status, $\chi^2$ (1, N = 266) = 1.42, p = .234. See Table 2.

**Table 2. Participants who Stopped Using an Online Service in the last 12 Months by Gender, Age Group and Disability Status**

| | Stopped using an online service | | | | | | |
|---|---|---|---|---|---|---|---|
| | Yes | | No | | Total | | *p* |
| | n | % | n | % | n | % | |
| Gender | | | | | | | .695 |
| Male | 81 | 62.8 | 48 | 37.2 | 129 | 100.0 | |
| Female | 95 | 65.1 | 51 | 34.9 | 146 | 100.0 | |
| Total | 176 | 64.0 | 99 | 36.0 | 275 | 100.0 | |
| Age group | | | | | | | .475 |
| 60-69 | 95 | 66.0 | 49 | 34.0 | 144 | 100.0 | |
| 70+ | 81 | 61.8 | 50 | 38.2 | 131 | 100.0 | |
| Total | 176 | 64.0 | 99 | 36.0 | 275 | 100.0 | |
| Disability status | | | | | | | .234 |
| With disability | 44 | 69.8 | 19 | 30.2 | 63 | 100.0 | |
| Without disability | 125 | 61.6 | 78 | 38.4 | 203 | 100.0 | |
| Total | 169 | 63.5 | 97 | 36.5 | 266 | 100.0 | |

In addition, Table 3 describes the results of the multiple-choice question regarding the security precautions participants take with their own devices. Of those who answered this question (n=274), the most common protective action was using a PIN or password followed by regular software updates. The use of two factor or multi-factor authentication, and unique passwords for every online service, were also common. Interestingly, the former, two/multi factor authentication, was about 11 percentage points lower than that of the average for the wider population.

**Table 3. Participant's most Common Online Security Actions**

| Older adults' protective actions* | n | % |
|---|---|---|
| Use a pin or password on any or all devices | 213 | 77.7 |
| Regularly update the software | 157 | 57.3 |
| Use two factor or multi factor authentication on any or all accounts | 118 | 43.1 |
| Use unique passwords for every online service | 117 | 42.7 |
| Regularly back up the content | 93 | 33.9 |
| Use a password manager | 62 | 22.6 |
| Use a VPN if connecting to the Internet using a Wi-Fi network you don't know | 48 | 17.5 |
| None of these | 10 | 3.6 |

Note: * n=274. Multiple-choice question.

On the other hand, when analyzing whether participants know where to report an online threat, the Chi-square test of independence found a significant association with gender, $\chi^2$ (1, N = 227) = 4.34, p = .037. The effect size for this finding, Cramer's V, was weak: .13. Male participants were more likely to say they know where to report an online risk than female participants. No significant association was found in terms of age, $\chi^2$ (1, N = 227) = 3.42, p = .065. Similarly, regarding disability status, the Chi-square test of independence did not reach significance, $\chi^2$ (1, N = 219) = 0.095, p = .758. See Table 4.

**Table 4. Participants' Knowledge about Where to Report Online Threat by Gender, Age Group and Disability Status**

| | Know where to report online threat | | | | | | |
|---|---|---|---|---|---|---|---|
| | Yes | | No | | Total | | *p* |
| | n | % | n | % | n | % | |
| Gender | | | | | | | .037 |
| Male | 50 | 46.3 | 58 | 53.7 | 108 | 100.0 | |
| Female | 39 | 32.8 | 80 | 67.2 | 119 | 100.0 | |
| Total | 89 | 39.2 | 138 | 60.8 | 227 | 100.0 | |
| Age group | | | | | | | .065 |
| 60-69 | 55 | 44.7 | 68 | 55.3 | 123 | 100.0 | |
| 70+ | 34 | 32.7 | 70 | 67.3 | 104 | 100.0 | |
| Total | 89 | 39.2 | 138 | 60.8 | 227 | 100.0 | |
| Disability status | | | | | | | .758 |
| With disability | 20 | 37.0 | 34 | 63.0 | 54 | 100.0 | |
| Without disability | 65 | 39.4 | 100 | 60.6 | 165 | 100.0 | |
| Total | 85 | 38.8 | 134 | 61.2 | 219 | 100.0 | |

On the other hand, it was found that the level of concern about three of the six online risks mentioned was associated with participants' gender (Table 5). As can be seen by the frequencies cross-tabulated in Table 6, concern about online extremism was associated with gender, $\chi^2$ (2, N = 269) = 9.44, p = .009. However, the effect size for this finding, Cramer's V, was weak: .18. Female participants were more likely to be extremely/very concerned about online extremism than male participants. Moreover, a significant but small association was found between concerns regarding misleading information and gender, $\chi^2$ (2, N = 268) = 8.17, p = .017, V = .17. Female participants were slightly more concerned about misleading information on the internet than male participants. When looking at the results for hate speech and gender, the association was also significant but small, $\chi^2$ (2, N = 269) = 7.49, p = .024, V = .16. In contrast no significant differences were found in terms of gender and the concerns about cyberbullying, $\chi^2$ (2, N = 271) = 5.95, p = .051, identify theft, $\chi^2$ (2, N = 275) = 2.88, p = .237, and conspiracy theories, $\chi^2$ (2, N = 269) = 0.281, p = .869.

**Table 5. Participants' Level of Concern about Different Online Risks by Gender**

| | | Gender | | | | Total | | *p* |
|---|---|---|---|---|---|---|---|---|
| | | Male | | Female | | | | |
| | | n | % | n | % | n | % | |
| Cyberbullying | Extremely/very concerned | 90 | 70.3 | 112 | 78.3 | 202 | 74.5 | .051 |
| | A little bit concerned | 21 | 16.4 | 24 | 16.8 | 45 | 16.6 | |
| | Not very/not at all concerned | 17 | 13.3 | 7 | 4.9 | 24 | 8.9 | |
| Online extremism | Extremely/very concerned | 87 | 68.0 | 108 | 76.6 | 195 | 72.5 | .009 |
| | A little bit concerned | 20 | 15.6 | 26 | 18.4 | 46 | 17.1 | |
| | Not very/not at all concerned | 21 | 16.4 | 7 | 5.0 | 28 | 10.4 | |
| Identity theft | Extremely/very concerned | 95 | 73.6 | 108 | 74.0 | 203 | 73.8 | .237 |
| | A little bit concerned | 24 | 18.6 | 33 | 22.6 | 57 | 20.7 | |
| | Not very/not at all concerned | 10 | 7.8 | 5 | 5 | 15 | 5.5 | |
| Misleading information | Extremely/very concerned | 85 | 67.5 | 99 | 69.7 | 184 | 68.7 | .017 |
| | A little bit concerned | 24 | 19.0 | 37 | 26.1 | 61 | 22.8 | |
| | Not very/not at all concerned | 17 | 13.5 | 6 | 4.2 | 23 | 8.6 | |
| Conspiracy theories | Extremely/very concerned | 83 | 64.8 | 95 | 67.4 | 178 | 66.2 | .869 |
| | A little bit concerned | 27 | 21.1 | 29 | 20.6 | 56 | 20.8 | |
| | Not very/not at all concerned | 18 | 14.1 | 17 | 12.1 | 35 | 13.0 | |
| Hate speech | Extremely/very concerned | 86 | 68.8 | 117 | 81.3 | 203 | 75.5 | .024 |
| | A little bit concerned | 24 | 19.2 | 21 | 14.6 | 45 | 16.7 | |
| | Not very/not at all concerned | 15 | 12.0 | 6 | 4.2 | 21 | 7.8 | |

The associations between levels of concern about each of the six online risks and age were also tested. The results show a significant but small association in terms of conspiracy theories and age, $\chi^2$ (2, N = 269) = 6.87, p = .032, V = .16. Participants aged 70 and over were more likely to be extremely/very concerned about conspiracy theories compared to those aged 60-69. Regarding hate speech and age, the association was significant, $\chi^2$ (2, N = 269) = 6.05, p = .049. The Cramer's V for this finding is a small .015. On the other hand, the differences between age and concerns about cyberbullying, $\chi^2$ (2, N = 271) = 0.723, p = .697, online extremism, $\chi^2$ (2, N = 269) = 4.86, p = .088, identify theft, $\chi^2$ (2, N = 275) = 5.19, p = .075, and misleading information, $\chi^2$ (2, N = 268) = 0.465, p = .793, did not reach statistical significance (Table 6).

**Table 6. Participants' Level of Concern about Different Online Risks by Age Group**

| | | Age group | | | | Total | | $p$ |
|---|---|---|---|---|---|---|---|---|
| | | 60-69 | | 70+ | | | | |
| | | n | % | n | % | n | % | |
| Cyberbullying | Extremely/very concerned | 103 | 73.0 | 99 | 76.2 | 202 | 74.5 | .697 |
| | A little bit concerned | 26 | 18.4 | 19 | 14.6 | 45 | 16.6 | |
| | Not very/not at all concerned | 12 | 8.5 | 12 | 9.2 | 24 | 8.9 | |
| Online extremism | Extremely/very concerned | 92 | 66.7 | 103 | 78.6 | 195 | 72.5 | .088 |
| | A little bit concerned | 29 | 21.0 | 17 | 13.0 | 46 | 17.1 | |
| | Not very/not at all concerned | 17 | 12.3 | 11 | 8.4 | 28 | 10.4 | |
| Identity theft | Extremely/very concerned | 106 | 73.6 | 97 | 74.0 | 203 | 73.8 | .075 |
| | A little bit concerned | 34 | 23.6 | 23 | 17.6 | 57 | 20.7 | |
| | Not very/not at all concerned | 4 | 2.8 | 11 | 8.4 | 15 | 5.5 | |
| Misleading information | Extremely/very concerned | 96 | 68.1 | 88 | 69.3 | 184 | 68.7 | .793 |
| | A little bit concerned | 34 | 24.1 | 27 | 21.3 | 61 | 22.8 | |
| | Not very/not at all concerned | 11 | 7.8 | 12 | 9.4 | 23 | 8.6 | |
| Conspiracy theories | Extremely/very concerned | 83 | 59.3 | 95 | 73.6 | 178 | 66.2 | .032 |
| | A little bit concerned | 37 | 26.4 | 19 | 14.7 | 56 | 20.8 | |
| | Not very/not at all concerned | 20 | 14.3 | 15 | 11.6 | 35 | 13.0 | |
| Hate speech | Extremely/very concerned | 97 | 69.3 | 106 | 82.2 | 203 | 75.5 | .049 |
| | A little bit concerned | 29 | 20.7 | 16 | 12.4 | 45 | 16.7 | |
| | Not very/not at all concerned | 14 | 10.0 | 7 | 5.4 | 21 | 7.8 | |

As can be seen in Table 7, the Chi-square test of independence showed no significant differences between the disability status of the participants and each of the six online risks. Disability was not associated with concerns about cyberbullying, $\chi^2$ (2, N = 263) = 2.20, p = .333, online extremism, $\chi^2$ (2, N = 261) = 4.96, p = .084, identity theft, $\chi^2$ (2, N = 266) = 0.267, p = .875, misleading information, $\chi^2$ (2, N = 261) = 0.325, p = .850, conspiracy theories, $\chi^2$ (2, N = 261) = 1.95, p = .376, and hate speech, $\chi^2$ (2, N = 260) = 0.477, p = .788.

**Table 7. Participants' Level of Concern about Different Online Risks by Disability Status**

| | | Disability status | | | | Total | | $p$ |
|---|---|---|---|---|---|---|---|---|
| | | With disability | | Without disability | | | | |
| | | n | % | n | % | n | % | |
| Cyberbullying | Extremely/very concerned | 50 | 80.6 | 145 | 72.1 | 195 | 74.1 | .333 |
| | A little bit concerned | 9 | 14.5 | 36 | 17.9 | 45 | 17.1 | |
| | Not very/not at all concerned | 3 | 4.8 | 20 | 10.0 | 23 | 8.7 | |
| Online extremism | Extremely/very concerned | 46 | 74.2 | 142 | 71.4 | 188 | 72.0 | .084 |
| | A little bit concerned | 6 | 9.7 | 39 | 19.6 | 45 | 17.2 | |
| | Not very/not at all concerned | 10 | 16.1 | 18 | 9.0 | 28 | 10.7 | |
| Identity theft | Extremely/very concerned | 48 | 76.2 | 148 | 72.9 | 196 | 73.7 | .875 |
| | A little bit concerned | 12 | 19.0 | 44 | 21.7 | 56 | 21.1 | |
| | Not very/not at all concerned | 3 | 4.8 | 11 | 5.4 | 14 | 5.3 | |
| Misleading information | Extremely/very concerned | 41 | 68.3 | 137 | 68.2 | 178 | 68.2 | .850 |
| | A little bit concerned | 13 | 21.7 | 48 | 23.9 | 61 | 23.4 | |
| | Not very/not at all concerned | 6 | 10.0 | 16 | 8.0 | 22 | 8.4 | |
| Conspiracy theories | Extremely/very concerned | 42 | 67.7 | 130 | 65.3 | 172 | 65.9 | .376 |
| | A little bit concerned | 15 | 24.2 | 40 | 20.1 | 55 | 21.1 | |
| | Not very/not at all concerned | 5 | 8.1 | 29 | 14.6 | 34 | 13.0 | |
| Hate speech | Extremely/very concerned | 48 | 78.7 | 148 | 74.4 | 196 | 75.4 | .788 |
| | A little bit concerned | 9 | 14.8 | 36 | 18.1 | 45 | 17.3 | |
| | Not very/not at all concerned | 4 | 6.6 | 15 | 7.5 | 19 | 7.3 | |

## 4. Discussion and Conclusions

The current study explored the attitudes and behaviors of older adults from New Zealand aged 60 years and older regarding different dimensions of cybersecurity and online safety. In addition to identifying general trends, the study also investigated whether each of these dimensions were associated with key demographics (i.e., gender, age group, and disability status). When looking at general trends, the results show that more than half of the older adults in our sample were concerned about the security of their personal information online. About 55% of older adults said they were extremely or very concerned while around 36% indicated being a little bit concerned about this matter. Similar trends have been reported in previous studies in the United States and New Zealand where most older adults indicated being apprehensive about how secure their personal information was on the internet with some indication that these concerns are on the rise (Boise et al., 2013; Kakulla, 2021; Pacheco, 2024). As unauthorized access to personal information, information misuse, and/or lack of transparency in information collection are among the most pressing issues for older adults regarding online privacy and security (Elueze & Quan-Haase, 2018; Knight et al., 2022), they might explain the rates reported not only here but also overseas. Similarly, our results show that most older adults are extremely/very concerned about each of the six online risks examined (i.e., cyberbullying, online extremism, identity theft, misleading information, conspiracy theories, and hate speech) with rates ranging from 66.2% to 75.5%.

Additionally, due to concerns regarding cybersecurity, the majority of older adults in our sample discontinued their usage of an internet-based service in the prior year. Specifically, 65% of participants reported doing so. An explanation for this result might be found in prior qualitative studies which indicate that limiting or avoiding the use of online tools or services is a common mitigation approach among older adults (Feng et al., 2023; Frik et al., 2019). When it comes to the most common security actions to prevent online threats, adults aged 60 years and older apply different approaches. In line with Huang and Bashir (2018), participants in our study reported adopting a combination of protective actions on their own devices, with the most common being the use of a PIN or password, regular software update, multi-factor authentication, and a unique password for every online service. However, while this evidence is relevant to understand older adults' behaviors, our results only describe technical measures such as the use of software and technical tools to prevent unwanted online experiences. Older adults often prioritize the use of social resources (i.e., family and close friends) to deal with cybersecurity and online safety threats (Nicholson et al., 2019). Future research should investigate how prevalent and effective these social resources are in older adults' coping strategies and support-seeking behaviors. Similarly, another finding surprisingly shows that most older adults do not know where to report potentially harmful or concerning online content. In this respect, only 4 in 10 participants indicated they know what websites, organizations, and/or agencies they can use or go to when encountering this sort of online content. It is argued that older adults who show lower levels of knowledge and confidence regarding addressing online threats exhibit a lesser degree of engagement in security-related behaviors in comparison to their younger counterparts (Jiang et al., 2016). Despite the fact that older adults may indeed partake in certain safety measures (e.g., creating secure passwords, regularly updating their devices), there may be other actions that they are less inclined to take, such as securing their devices (Branley-Bell et al., 2022). The findings from the present study present a similar trend, indicating that the online safety and security behaviors of older adults constitute a complex issue that warrants further exploration.

From the general trends described above, it is evident that older adults express concern regarding risks on the internet. They are actively implementing certain measures to safeguard themselves against online threats; however, they encounter obstacles such as unfamiliarity with the appropriate channels to report potentially harmful online content. The empirical evidence derived from the present study demonstrates that older adults aged 60 years and above require prompt and focused assistance to enhance their confidence and skills, in order to make their online experience safer. Provision of assistance is imperative due to older adults' increasing adoption of digital technologies, and the importance these tools gained among the older population, particularly since the COVID-19 pandemic. So far, interventions to prevent or mitigate online threats for older adults have focused on education and training programs as well as awareness raising. While evaluation of their efficacy is a pending task, online safety and cybersecurity interventions have traditionally overlooked the input of older adults during the development phase. In accordance with the suggestion made by Mannheim et al. (2019) regarding the involvement of older adults in technology design, a similar approach is required when creating support systems for older adults in the realm of online safety and cybersecurity. Including the voice of older adults in the design stage of support interventions can contribute a more profound comprehension of the challenges that affect them, their specific needs, and the appropriate measures that are more pertinent to their situation. Moreover, this approach can help create greater impact and effectiveness of these interventions (Mannheim et al., 2019).

On the other hand, regarding the role of key demographics, the current study has found some statistical differences suggesting older adults' heterogeneity regarding cybersecurity and online safety attitudes and behaviors. For instance, in terms of gender, older males (46.3%) were much more likely to say they know where to report potentially harmful content than older females (32.8%). Gender differences were also found regarding the levels of concern about online risks. In this respect, compared to older males, older females reported higher rates of concerns about hate speech, online extremism, and misleading information. Evidence about the link between gender and cybersecurity and online safety has traditionally centered on the general adult population. Despite some exceptions (Branley-Bell et al., 2022), this evidence has mostly found gender differences with females being more likely to have lower digital security skills and awareness than males (Alotaibi & Alshehri, 2020; Dodel & Mesch, 2018; McGill & Thompson, 2021). The analysis undertaken here about the role of gender provides a closer look at the cybersecurity and online safety attitudes and behaviors among older adults aged 60 years and older.

Similar to the preceding, studies on age differences in cybersecurity and online safety have primarily focused on the broader adult population (Branley-Bell et al., 2022; Dodel & Mesch, 2018). Nevertheless, the present study has made a noteworthy discovery: within the older adult cohort, there exist significant variations in age. In this respect, regarding online risks, the data uncovered higher levels of concern about conspiracy theories and hate speech among older adults aged 70 years and above compared to their 60–69-year-old counterparts. The differences were of 14 and 13 percentage points, respectively. Older adults aged 70 years and older were also slightly more concerned about the security of their personal information online than those aged between 60 and 69 years. It has been documented that older adults place more importance on the privacy and security of their personal information online than younger age groups (Kolimi et al., 2012). However, prior to the current study, evidence about how these attitudes differed within the older adult cohort was lacking.

The present study has argued about the inclusion of older adults in the design of online safety and security programs. However, it is important to note that these support interventions should also acknowledge the fact that the older adult group is diverse in nature. It is surprising that despite the existence of online safety and cybersecurity interventions for a considerable time, the heterogeneity of older adults regarding experiences of online risks and threats has gone unnoticed. The findings of the present study complement the existing evidence regarding the varying online behaviors and attitudes of older adults in terms of internet access (Hargittai & Dobransky, 2017; van Deursen & Helsper, 2015). Recognizing the heterogeneity of older adults in terms of their online safety and cybersecurity attitudes and behaviors will aid in the development of interventions that are accessible and inclusive, cater to the needs of the target population, and are suitable for their intended purpose.

Finally, none of the dependent variables examined in the present study showed an association with disability status. While this relationship has not been well researched yet, the limited available evidence suggests that older adults with moderate cognitive impairment are more sensitive to online threats (Mentis et al., 2019). Our findings are somewhat surprising considering that older adults are more sensitive to online risks as their functioning declines (Lichtenberg et al., 2016). Further research on the role of disability in older adults' internet safety and security habits and attitudes would allow us to develop a higher level of accuracy in this matter.

### 4.1. Limitations

There were some limitations to the present study which future research could address. While this study managed to find some statistical associations between categorical variables, it does not mean these links explain causation. Furthermore, the probability of social desirability bias in participants' responses is another potential limitation. Due to the sensitive nature of the research topic, some participants might not have disclosed their personal experiences and views regarding online safety and security. Finally, as previously mentioned, we conducted secondary data analysis of the New Zealand's Internet insights Survey. Thus, another limitation is that our analysis and interpretation of the data was restricted to the original design of the survey tool. For instance, while the survey gathered evidence about how concerned participants are about online risks, it did not ask participants, whether they were personally targeted by any of them, how frequently they happened, and what was the emotional impact, if any. Gathering data about these and other variables in subsequent research will improve the understanding of older adults' online safety and security behaviors and attitudes.

## Acknowledgments

The author would like to thank InternetNZ (https://internetnz.nz/) for providing access to the data underpinning this paper.